**Phonon Gravity, Non-equilibrium QFT, and the Tolman Thermal Equivalence Principle**

A. Jourjine

FG CTP, Hofmann Str. 8-12, 01281 Dresden, Germany

**Abstract**

We describe an extension of the Keldysh method for fermions from constant temperature to steady state case with spatially varying temperature field. This is done with the use on the imaginary section of the Keldysh path of a thermal Hamiltonian obtained from the zero temperature relativistic Hamiltonian by coupling its density multiplicatively to the temperature field. We show that the two Hamiltonians commute, provided appropriate boundary conditions are imposed. A microscopical equation on the temperature field and the corresponding microscopical Fourier law of heat transfer are derived for the non-relativistic case. We discuss application of the proposed method to the thermoelectric effect and point out a remarkable correspondence between the non-equilibrium thermal Hamiltonian and the zero temperature fermionic Hamiltonian in general relativity for the metric obtained from the Minkowski metric by rescaling time with the inverse temperature. Our results suggest the existence of the correspondence principle between gravitating equilibrium and non-gravitating non-equilibrium quantum field theories, which we call the Tolman thermal equivalence principle in honor of his pioneering work.

1. **Introduction**

The Keldysh closed time path perturbation theory is one of the most significant advances in the non-equilibrium quantum field theory [1, 2]. One drawback of the method is that it is defined for constant temperatures. This is a limiting factor to its applications in the areas, such as thermoelectric effect, where the interesting physics comes from spatial variations of temperature.

The microscopic foundations of the Fourier law of heat transfer have been a subject of research for some time. Since the review of the state of the art in [3] a number of further developments took place using various explicit models and Boltzmann equation approximations in both semi-classical and quantum-mechanical setting in an attempt to provide the Fourier law with a solid microscopic foundation, to explore deviations from it, and to set up the experimental conditions for micro-measurements of temperature distributions in materials [4 - 16]. This work has become all the more significant, because temperature variations can play significant role in the work of mesoscopic devices, such as nano-junctions. Typically the Fourier law is derived first. Conservation of energy is then may be used to derive the heat equation, from which a steady state equation on temperature field is deduced.

Recently, there appeared a number of publications [17 - 20] describing application of the Keldysh method to the density functional theory, where a temperature field is introduced that couples multiplicatively to the energy density. One thing lacking in this work is a method to determine the temperature field itself that is based on the first principles of quantum field theory. From non-equilibrium thermodynamics we know that the temperature field is governed by the heat equation, which reduces to Laplace-like equation for the steady state case. However, the derivation of the heat equation is based on the Fourier's law, whose microscopical justification is rather subtle. A quantum-statistical derivation of the temperature equation from the first principles of quantum field theory does not seem to exist. Such a derivation is therefore desirable in view of many practical applications of the thermoelectrical effect and other phenomena, where temperature gradients play an essential role. Another area of application could lie in the theory of evolution of stars, where temperature gradients are essential to understand, especially in the regime when quantum relativistic effects become significant [28]. One possibility for such a derivation lies in the use of super-statistics, where the statistical sum for steady state includes an additional statistical averaging over all temperature distributions [29]. However, the relation of the formalism to a conceptually simple equation on temperature field is not very transparent.

In this paper, following the pioneering suggestion of Lüttinger [21] to use a fictitious gravitational field, we derive the steady state heat equation for fermions using a non-relativistic approximation of the thermal Hamiltonian, obtained by coupling multiplicatively the relativistic Hamiltonian density at zero temperature with the inverse temperature field. As a consequence, using the coupling we obtain the Fourier law of heat transfer from the steady state heat equation.

Finally, we point out a remarkable correspondence between the non-equilibrium thermal fermionic Hamiltonian in flat space-time and the equilibrium general relativistic fermionic Hamiltonian for metric line element $ds^2 = \beta^2(\vec{x}) dt^2 - d\vec{x}^2$, where $\beta(\vec{x})$ is the dimensionless inverse temperature field. The correspondence is suggested by the work of Tolman in general relativity on the weight of heat [22]. We speculate about the existence of a general thermal equivalence principle that reflects a correspondence between non-equilibrium steady state quantum field theory for Minkowski space-time for temperature field $\beta(\vec{x})$ and the equilibrium relativistic quantum field theory for $\beta(\vec{x})$.

In the next Section we describe the thermal Hamiltonian and its properties, while in the third Section we consider application of the Keldysh method to the thermoelectric effect and derive a microscopical Laplace-like equation for the temperature field. At the end of this section we discuss the proposed Tolman thermal equivalence principle.

## 2. The Thermal Hamiltonian

The Keldysh method allows the construction of a consistent field-theoretic perturbation theory of the statistical averages of the *T*-product of fields in a heat bath at constant temperature, which are the basic elements of quantum statistics of the systems with an infinite number of the degrees of freedom. Given an equilibrium temperature $T_0$ the averages, i.e., the Green functions, are given by

$$G_{\beta_0}(x,x',\ldots) = \langle T_C[P(\psi(x),\psi(x'),\ldots)U\,]\rangle_{\beta_0}, \quad U = \exp\left(-\frac{i}{\hbar}\int d^3x\,\mathcal{H}(x)\right), \qquad (1)$$

where $P(\psi(x),\psi(x'),\ldots)$ is a polynomial of fermionic fields $\psi(x), \psi(x'),\ldots$, $\langle A\rangle_{\beta_0} = Z_0^{-1}tr(A\rho_0)$ is the statistical average with the density matrix $\rho_0 = \rho(t_i)$ at the initial time $t_i$, the partition function $Z = tr\rho$ is evaluated at $t_i$, and $T_C$ is the time ordered product along contour *C* in the complex plane that runs from the initial time $t_i = 0$ to the final time $t_f = +\infty$, then back to $t_i$ and then from $t_i$ to $t_i - i\beta_0$, $\beta_0 = T_0^{-1}$. The fields $\psi(x), \overline{\psi}(x)$ satisfy the canonical equal time anticommutation relation $\{\psi_\alpha(t,\vec{x}),\psi_\beta^{+}(t,\vec{y})\} = \delta_{\alpha\beta}\delta(\vec{x}-\vec{y})$. In the interaction picture the Hamiltonian density $\mathcal{H}(x)$ in (1) is replaced by its interaction part $\mathcal{H}_I(x)$, defined using free fields.

We now extend the Keldysh method from the thermal equilibrium to the thermal steady state setting by replacing *H* on the imaginary section of the path *C* with the thermal Hamiltonian $H_\beta$, where

$$H_\beta = \int d^3x\, \beta(x)\mathcal{H}(x), \qquad (2)$$

where $\beta(x) = T_0/T(x)$ is the dimensionless temperature field while $T_0$ is some arbitrary reference temperature. Such a replacement is intuitively appealing.

We now note that the statistical averaging in (1) uses the zero temperature eigenstates states of the Hamiltonian and averages them with statistical weights that depend on temperature. These statistical weights are the eigenvalues of $\exp(-\beta_0 H)$. Therefore, it is natural to assume that the eigenstates of $H_\beta$ are the same as for $H$ and only the eigenvalues of $\exp(-\beta_0 H_\beta)$ may change. This means that the commutator of the two Hamiltonians must vanish

$$[H, H_\beta] = 0. \qquad (3)$$

Before we consider the consequences of (3) a few remarks are in order. First, to construct $H_\beta$ we have to work within the framework of the second quantization. Only in this context we can speak of the Hamiltonian density $\mathcal{H}(x)$. Secondly, we will assume here that all Hamiltonians are derived from the corresponding general-relativistic symmetric energy-momentum tensors, thus making them unique, because in general relativity energy density is an observable.

Let us consider (3) in detail. For the relativistic case with the Hamiltonian

$$\mathcal{H}(x) = \psi^+\left[\left(-\frac{i}{2}\hbar c\vec{\partial} - e\vec{A}\right) + eA_0 + \gamma^0 mc^2\right]\psi \equiv \psi^+ \hat{H}\psi, \qquad (4)$$

where $\vec{\partial} \equiv \alpha^k(\vec{\partial}_k - \overleftarrow{\partial}_k)$, $\vec{A} = A_k \alpha^k$, $\alpha^k = \gamma^0\gamma^k$, $k = 1, 2, 3$, $\{\gamma^\mu, \gamma^\nu\} = 2\,diag(1,-1,-1,-1)$, and $\hat{H}$ is the one-particle Hamiltonian operator, we obtain

$$[H, H_\beta] = 2\int d^3x\, \partial_k \beta J^k, \qquad (5)$$

$$J^k = \psi^+\{\alpha^k, \hat{H}\}\psi. \qquad (6)$$

It follows from (4) that the current $J^k$ can be decomposed into four components

$$\vec{J} = \vec{J}_1 + \vec{J}_2 + \vec{J}_3 + \vec{J}_4 \qquad (7)$$

where $\vec{J}_1 = \hbar c \psi^+ \vec{\partial} \psi$, is the particle current, $\vec{J}_2 = (\hbar c/2) \psi^+ [\vec{\alpha}, \alpha^k] \vec{\partial}_k \psi$ is the spin density current, $\vec{J}_3 = 2e(\psi^+ \psi) \vec{A}$ is proportional to the charge density and $\vec{J}_4 = 2e(\psi^+ \vec{\alpha} \psi) A_0$ is proportional to the charge density current. The three-currents $\vec{J}_p$, $p = 1, 2, 4$ in (6) are spatial components of the four-currents $J^\mu{}_p$ that are conserved on-shell: in the steady state case when currents do not depend on time $\partial_0 J^0{}_p = 0$ and, therefore, are divergence free: $\partial_k J^k{}_p = 0$. Further, the gauge field $A_\mu = (A_0, A_k)$ for our steady state system should be considered in the physical (the Lorentz) gauge with $\partial_k A_k = \partial_0 A_0 = 0$. Having this in mind we can transform (5) into a surface integral so that $[H, H_\beta] = 2 \int d^2 S_k (\beta J^k)$, where $d^2 S_k$ is the surface element. We conclude that the commutator for the total current (6) vanishes if the following conditions are satisfied

$$\int d^2 S_k (\beta J^k) = 0, \qquad \vec{E} \cdot (\psi^+ \vec{\alpha} \psi) = 0, \ \vec{E} = -\vec{\partial} A_0. \qquad (8)$$

These conditions, when averaged, may be considered as a part of the definition of the steady state, in addition to vanishing of time derivatives of the appropriate averaged observables. The first condition in (8) is a boundary condition, while the second simply means that in steady state the electric field may not pump energy into the system.

The boundary conditions in (8) also have an apparent physical meaning: in order for steady state to be well-defined the total average temperature-weighted influx into the system of particles, spin, and charge must vanish. Obviously, with an appropriate definition of some of the currents the steady state definition would also apply to the non-relativistic case. Note that our derivation applies only to the second-quantized systems. The condition (3) cannot be interpreted as a constraint on quantum-mechanical operators, because it easy to see that then only constant temperatures are allowed [23]. Note that (3) is also satisfied for the grand canonical ensemble, where for particle density $\mathcal{N}$ and chemical potential we replace $\mathcal{H} \to \mathcal{H} - \mu \mathcal{N}$.

## 3. The Steady State Keldysh Method and the Thermoelectric Effect

As an example of the application of the Keldysh method we will consider the thermoelectric effect for electron gas in metals. First, we briefly describe the classical quantum mechanical derivation of the properties of the thermoelectric effect. The key is the use of the Boltzmann equation with a relaxation time approximation. This is supplemented with the use of the modified Fermi distribution function for the electron gas as the unperturbed distribution. One obtains a reasonably good approximation for the heat and the charge currents. However, to carry the calculations through one has to use a slight of hand. Since quantum statistics of point particles cannot describe steady state non-equilibrium systems, when applying the Boltzmann equation one replaces the Fermi distribution at constant temperature with the one where temperature becomes dependent on spatial coordinates [24]. The replacement is called the use of the local equilibrium approximation.

However, while Fermi distribution comes from the statistical sum, the local equilibrium distribution function originates from a different source: it is the Wigner transform of the two-point correlation function derived via the Keldysh method. This distinction is important for field-theoretic description of the thermoelectric effect.

We now outline the path from (1) to the Boltzmann equation and its solution that corresponds to the local equilibrium approximation. Details can be found in [1, 2]. According to Keldysh the properties of the system are described by a matrix of four Greens functions

$$G(x,x') = \begin{bmatrix} G^c(x,x') & G^-(x,x') \\ G^+(x,x') & \tilde{G}^c(x,x') \end{bmatrix}, \qquad x = (t, \vec{x}). \tag{9}$$

Functions $G^c, \tilde{G}^c$ describe the dynamical properties of the system, while functions $G^\pm$ describe its statistical properties. The functions are related linearly to the systems advanced and retarded Green functions $G^a = G^c - G^+$, $G^r = G^c - G^-$ and to the one-particle density matrix $F = G^+ + G^-$. This can be seen from their definitions and from solutions of the equations that the functions satisfy in the free field case. They are defined by

$$G^c(x,x') = -i\langle T_C[\psi_0(\vec{x},t_+)\psi^+{}_0(\vec{x}',t'_+)U]\rangle_\beta$$
$$\tilde{G}^c(x,x') = -i\langle T_C[\psi_0(\vec{x},t_-)\psi^+{}_0(\vec{x}',t'_-)U]\rangle_\beta \tag{10}$$
$$G^\pm(x,x') = -i\langle T_C[\psi_0(\vec{x},t_\pm)\psi^+{}_0(\vec{x}',t'_\mp)U]\rangle_\beta,$$

where argument $t_\pm$ signifies the upper/lower horizontal branch of the Keldysh path.

Of most interest for us is the function $G^+$, because it can be shown that in certain approximation it satisfies the Boltzmann equation. For free field translation invariant relativistic case in the momentum space it is given by

$$G^+(\vec{p}, p^0) = 2\pi i n_p \delta(p^2 - m^2), \qquad (11)$$

where $n_p$ is the Fermi distribution with chemical potential $\mu$

$$n_p = [1 + \exp \beta(p^0 - \mu)]^{-1}. \qquad (12)$$

In the non-relativistic case one replaces $\delta(p^2 - m^2)$ using appropriate particle energy $\varepsilon_p$ according to $\delta(p^2 - m^2) \to \delta(p^0 - \varepsilon_p)$.

When interactions are present $G$ in (9) satisfies the matrix Dyson equation, given in its differential form by

$$(i\partial_t - \hat{H}(x))G(x, y) = i\int dx' \Sigma(x, x')G(x', y) + \delta(x - y), \qquad (13)$$

where $\hat{H}(x)$ is the one-particle Hamiltonian operator (4) and $\Sigma(x, x')$ is the self-energy matrix defined by summing of the contributions to $G$ from all diagrams with one incoming and one outgoing line. It components are denoted by

$$\Sigma = \begin{bmatrix} \Sigma^c & \Sigma^- \\ \Sigma^+ & \tilde{\Sigma}^c \end{bmatrix}. \qquad (14)$$

Construction of self-energies for steady state system via perturbation theory that gives positive spectral function for equilibrium systems was discussed in [27].

In order to obtain the Boltzmann equation we now use the Wigner transform of $G^+(x, y)$. The Wigner transform and its inverse are given by

$$W^+(x, p) = \int d^4 y \, G^+(x + y/2, x - y/2)\exp(ipy),$$

$$\qquad (15)$$

$$G^+(x, y) = \frac{1}{(2\pi)^4} \int d^4 y \, W^+((x + y)/2, p)\exp(-ip(x - y)).$$

From (15) we obtain the distribution function $f(\vec{x}, \vec{p}, t)$ by integration over all energy states

$$f(\vec{x}, \vec{p}, t) = \frac{1}{2\pi i} \int dp^0 W^+(x, p). \tag{16}$$

It satisfies the Boltzmann equation when external fields are quasi-classical and interaction is small

$$\left[\partial_t + \vec{v} \cdot \vec{\partial}_x + e\left(\vec{E} + c^{-1}\vec{v} \times \vec{B}\right) \cdot \vec{\partial}_p\right] f(\vec{x}, \vec{p}, t)$$

$$= \frac{1}{2\pi i} \int dp^0 \left[\Sigma^+(x, p) G^-(x, p) - \Sigma^-(x, p) G^+(x, p)\right]. \tag{17}$$

The distribution $f(\vec{x}, \vec{p}, t)$ is the main ingredient in the computation of the charge and heat currents for the thermoelectric effect. Its steady state approximation is the local equilibrium condition

$$f(\vec{x}, \vec{p}, t) = \left[1 + \exp \beta(\vec{x})(\varepsilon_p - \mu)\right]^{-1}. \tag{18}$$

From $f(\vec{x}, \vec{p}, t)$ in (18) we can reconstruct $G^\pm(x, y)$ that enter (17), provided we know the dependence of $W^+(x, p)$ on $p^0$, or if it is not known up to the terms lost in integration over energy in (16). The only unknown entity in (17) is $\beta(\vec{x})$. Once it is known we can compute the heat and charge transport equations as a series of approximations as a power series in the coupling constant and the Fourier transform $\beta(\vec{k})$. For example, in the quantum-mechanical setting the first order correction to the zero-temperature eigenvalues $E_k^0$ will be $E_k^1 \propto E_k^0 \beta(\vec{k})$.

We now turn to the determination of the temperature field itself. First we will derive an equation for the non-relativistic case under the assumption that the thermal Hamiltonian is given by (2). Clearly we cannot derive a non-trivial equation on the temperature field in the relativistic case. The reason for this is that relativistic coupling of temperature to the Hamiltonian is linear. Only in the non-relativistic setting the coupling results in a meaningful equation. In order to derive a relativistic equation on temperature we have to modify the thermal Hamiltonian by adding to (2) terms that nonlinearly depend on temperature gradients. Since heat conduction in solids is carried out by phonons, clearly additional terms in (2) must come from phonon contributions. The necessary modification will be discussed below after we dealt with the practically important non-relativistic case.

To reduce $H_\beta$ in (2) to its non-relativistic approximation we need to decouple the light, the upper two components of the spinors in $H_\beta$, from the heavy lower two components. To do this we can use the Foldy-Wouthuysen transformation [25]. Using in addition a gauge transformation to get rid of time-dependent terms, we obtain the desired thermal Hamiltonian

$$H_{th} = \int d^3 x \psi^+ \hat{H}_{th} \psi,$$

$$\hat{H}_{th} = \beta \left[ \left( mc^2 + eA^0 + e\phi \right) + \frac{1}{2m} \left( -i\hbar \vec{\partial} - \frac{e}{c} \vec{A} - \frac{i}{2} \hbar \vec{\partial} \ln \beta \right)^2 - \frac{e\hbar}{2mc} \vec{\sigma} \cdot \vec{B} \right],$$

(19)

where $\hat{H}_{th}$ is the one-particle Hamiltonian with terms up to the first order in $(mc)^{-1}$, $\vec{B}$ is static magnetic field, and we separated the electrical field potential into the external potential $A^0$ and the electron self-interacting part

$$\phi = \frac{1}{2} \int d^3 y d^3 y' \psi^+(y) |\vec{y} - \vec{y}'|^{-1} \psi(y). \tag{20}$$

Note the contribution of the rest energy of the electron. It is absent when temperature is constant. This leads to non-trivial physical changes. The thermal Hamiltonian has a number of interesting features. To start with, it is gauge-invariant with respect to the modified (thermal) gauge transformations

$$A^k \to A^k - \partial_k \lambda, \qquad A^0 \to A^0 + (\beta_n c)^{-1} \partial_t \lambda. \tag{21}$$

Further, it has an explicit non-linear dependence on the temperature field, although on the relativistic level the temperature coupling is linear. The dependence allows us to derive the needed equation for the temperature field by demanding that for given boundary conditions the average of the thermal Hamiltonian $\langle H_{th} \rangle_\beta$ attains its maximum value. Clearly, the extremum must be a maximum, for temperature must be positive. In other words, it is such that the variation of $\langle H_{th} \rangle_\beta$ with respect to $\beta(\vec{x})$ vanishes.

$$\frac{\delta \langle H_{th} \rangle_\beta}{\delta \beta(\vec{x})} = 0. \tag{22}$$

Note the use of averaging. It is applied to avoid an over-constrained operator equation in Fock space. As a result of the averaging, the equation on the temperature field is the second order partial differential equation. Since we assumed that the dynamics of fields is derived at zero temperature, the variation with respect to $\beta(\vec{x})$ must be carried assuming that all fields have a fixed value.

For simplicity consider the case without magnetic field. We obtain the reduced thermal Hamiltonian

$$H_{th} = \int d^3x\, \psi^+ \left[ -\frac{\beta \hbar^2}{2m} \left( \vec{\partial} + \frac{1}{2} \vec{\partial} \ln \beta \right)^2 + \beta V \right] \psi, \tag{23}$$

where $V = (mc^2 + eA^0 + e\phi)$. After elimination of terms that are linear in $\vec{\partial}\beta$, ignoring the resulting surface terms, and defining $\chi = \ln \beta$ we obtain a simpler expression

$$H_{th} = \int d^3x\, e^\chi \left[ \frac{\hbar^2}{8m} \psi^+ \psi \left( \vec{\partial}\chi \right)^2 - \frac{\hbar^2}{4m} \left( \vec{\partial}^2\psi^+ \psi + \psi^+ \vec{\partial}^2\psi \right) + \psi^+\psi V \right] \tag{24}$$

After averaging we obtain from (24) the final form of the equation on the temperature field in absence of magnetic fields

$$\vec{\partial}^2 \chi + e^\chi \frac{\vec{\partial}\langle \psi^+\psi \rangle_\beta}{\langle \psi^+\psi \rangle_\beta} \cdot \vec{\partial}\chi + e^\chi \left[ 2 \frac{\langle \vec{\partial}^2\psi^+\psi + \psi^+\vec{\partial}^2\psi \rangle_\beta}{\langle \psi^+\psi \rangle_\beta} - \frac{8m}{\hbar^2} \langle V \rangle_\beta \right] = 0. \tag{25}$$

For small temperature deviations with $\ln(1+\theta) \approx \theta$ we obtain the linearized form of (25)

$$\vec{\partial}^2 \theta + \frac{\vec{\partial}\langle \psi^+\psi \rangle_\theta}{\langle \psi^+\psi \rangle_\theta} \cdot \vec{\partial}\theta + \left[ 2 \frac{\langle \vec{\partial}^2\psi^+\psi + \psi^+\vec{\partial}^2\psi \rangle_\theta}{\langle \psi^+\psi \rangle_\theta} - \frac{8m}{\hbar^2} \langle V \rangle_\theta \right] = 0, \tag{26}$$

where the derivatives of $\theta$ are not assumed to be small and the averaging is now at a constant reference temperature $T_0$. We see that the source of the temperature Laplacian is composed from two contributions: one from average normalized kinetic energy and one from the potential energy, including contributions from an external field. The linear derivative term that determines decay or growth of excitations is determined by the normalized particle density divergence.

We emphasize that equation (25, 26) can be derived only for the non-relativistic case. The equation on temperature field obtained from the relativistic thermal Hamiltonian (2) is trivial. To obtain non-trivial equation for steady state relativistic case we have to realize that in our derivations we omitted the contribution from the phonons. Triviality of the heat equation derived from (2) indicates that in the relativistic case phonon contribution to the overall thermal Hamiltonian of the system cannot be ignored.

We now show how the modification of (2) by adding the energy density of the phonons can be done in an essentially unique way. The price for the uniqueness is the appearance of phenomenological quantities that depend on the detailed structure of the system in question.

Consider massless scalar phonon field $\vec{\Phi}(x)$ defined in a system, which for simplicity we will assume to be isotropic. The hint for the simplest modification is comes from thermodynamics. We know that the heat equation for the steady state is the second order differential equation. Considering the $O(3)$ symmetry of the system, the following Hamiltonian for the phonons appears to be natural

$$H_\Phi = \frac{\lambda}{2} \int d^3x \, \vec{\partial}\beta(x) \cdot \vec{\partial}\beta(x), \tag{27}$$

where we identify

$$\beta(x) = f(\vec{\Phi}^2(x)) \tag{28}$$

and $f(\Phi)$, $\lambda$ is some function, we can call it the thermal function of sate, and a constant coefficient that must be determined from the detailed dynamics of the phonon field. Since we are only interested in the equation on the temperature field the detailed nature of function $f(\Phi)$ is not important for this discussion. The coefficient $\lambda$ is also not known beforehand, since it is a purely phenomenological quantity. However, its influence on the dynamics of temperature fields is simply parametric. This should be sufficient for many applications. Having all this in mind, we obtain the full thermal Hamiltonian

$$H_t = H_\Phi + H_\beta,\tag{29}$$

$$H_\Phi = \frac{\lambda}{2}\int d^3x\, \vec{\partial}\beta(x)\cdot\vec{\partial}\beta(x),\quad H_\beta = \int d^3x\, \beta(x)\mathcal{H}(x),$$

where $\mathcal{H}(x)$ is the relativistic zero temperature fermion Hamiltonian density. From (29) we obtain the relativistic equation on temperature field

$$\lambda\vec{\partial}^2\beta = \mathcal{H}(x),\tag{30}$$

and its non-relativistic approximation

$$(\lambda+1)\vec{\partial}^2\theta + \frac{\vec{\partial}\langle\psi^+\psi\rangle_\theta}{\langle\psi^+\psi\rangle_\theta}\cdot\vec{\partial}\theta + \left[2\frac{\langle\vec{\partial}^2\psi^+\psi + \psi^+\vec{\partial}^2\psi\rangle_\theta}{\langle\psi^+\psi\rangle_\theta} - \frac{8m}{\hbar^2}\langle V\rangle_\theta\right] = 0,\tag{31}$$

From (31) we can derive the Fourier law of heat transfer in its microscopical form. For a given temperature field $T(x)$, the macroscopical Fourier law defines the heat transfer current by

$$\vec{Q} = -\kappa\,\vec{\partial}T(x),\tag{32}$$

where $\kappa = \kappa(x)$ is the heat conductivity parameter. For the macroscopic Fourier law it is a phenomenological parameter determined experimentally. The heat equation follows from conservation of energy and is given by

$$c_V\frac{\partial T}{\partial t} = \vec{\partial}\cdot\left(\kappa\,\vec{\partial}T(x)\right),\tag{33}$$

where $c_V$ is the specific heat parameter. Unlike $\kappa$ the specific heat $c_V$ can be determined microscopically if the statistical sum of the system is known. We can now extract the microscopical steady state version of (32) from the microscopic steady state heat equation (31). For small temperature deviations with $\ln(1+\theta) \approx \theta$ we obtain the linearized steady state heat equation

$$\vec{\partial}\left(\left(\lambda + \frac{1}{2}\frac{\hbar^2}{4m}\langle\psi^+\psi\rangle_\theta\right)\cdot\vec{\partial}\theta\right) = E, \tag{34}$$

$$E = \left[-\frac{\hbar^2}{4m}\langle\vec{\partial}^2\psi^+\psi + \psi^+\vec{\partial}^2\psi\rangle_\theta + \langle V\rangle_\theta\langle\psi^+\psi\rangle_\theta\right], \tag{35}$$

where $\langle\ldots\rangle_\theta$ is the quantum statistical average calculated using the Keldysh method. Equation (34) is an integro-differential equation on the temperature field. We conclude that the relativistic multiplicative coupling of temperature field to fermion Hamiltonian density results in the non-relativistic approximation and for small temperature variations in the steady state heat equation that is essentially the Laplace equation with the source given by the average energy of the system. Given the boundary conditions on the temperature, solutions of (34, 35) exist and are unique.

In the first approximation we can ignore the functional dependence of (34, 35) on the temperature field. Then (34, 35) turns into an inhomogeneous second order differential equation. Any solution of such an equation can be represented as a sum of solutions for inhomogeneous and homogenous equation with $E = 0$. If we can neglect the contribution of the inhomogeneous term in comparison with the homogeneous term, the solution of (34) is determined by the boundary conditions alone and does not depend on the internal distribution of energy. For the steady state case with $E = 0$ we now can extract from (34) the microscopical steady state Fourier law

$$\vec{Q} = -\kappa\cdot\vec{\partial}\theta, \tag{36}$$

where the heat conductivity is given by the quantum-statistical average of the electron density operator

$$\kappa = \lambda + \frac{1}{2}\frac{\hbar^2}{4m}\langle\psi^+\psi\rangle_\theta. \tag{37}$$

Having derived the microscopical heat equation, we now would like to point out a remarkable correspondence between the *non-equilibrium* quantum fermionic steady state system we considered, where the dynamics is defined on flat space-time, and the *equilibrium* fermionic systems on a curved space-time with static gravitational field with metric given by line element

$$ds^2 = c^2 V^2(\vec{x}) dt^2 + d\vec{x} \cdot d\vec{x}. \tag{38}$$

The Hamiltonian of the corresponding relativistic Schrödinger equation is given by [26]

$$\hat{H}_g = \gamma^0 mc^2 V + e\Phi + \frac{c}{2}\vec{\alpha} \cdot \left[ \left(-i\hbar\vec{\partial} - (e/c)\vec{A}\right) V + V\left(-i\hbar\vec{\partial} - (e/c)\vec{A}\right) \right]. \tag{39}$$

We note that with replacements $\beta(\vec{x}) \to V(\vec{x})$, $\beta A^0 \to \Phi$, $\vec{A} \to \vec{A}$ the Hamiltonians (4) and (39) are identical. This remarkable correspondence is in accordance with the fact [22] that in general relativity with static gravitational field for systems in thermodynamical equilibrium the equilibrium temperature $T$ is not a constant but is the Tolman temperature $T_T = T_0/\sqrt{g_{00}(\vec{x})}$, $T_0 = const$, where $g_{00}(\vec{x})$ is the zero-zero component of the metric tensor $g_{\mu\nu}$. In our case in (39) we have $g_{00}(\vec{x}) = cV$. Therefore, the corresponding equilibrium metric is obtained by setting $V(\vec{x}) = \beta(\vec{x})$, which results in

$$ds^2 = c^2 \beta^2(\vec{x}) dt^2 + d\vec{x}^2. \tag{40}$$

We further note that the phonon contribution to the total thermal Hamiltonian can be also interpreted in terms of general relativity. If we denote by $R = \int d^3x \sqrt{g}\, \mathcal{R}(x)$ the integrated curvature scalar for metric (40) then we obtain that

$$R = \int d^3x \sqrt{g}\, \mathcal{R}(x) = \int d^3x \frac{1}{2}\left(\vec{\partial}\beta\right)^2, \tag{41}$$

Therefore, the total thermal Hamiltonian (29) that includes the phonon contribution can be written as

$$H_t = \lambda R + \hat{H}_g. \tag{42}$$

This is the Hamiltonian for a quantum spinor field interaction with a classical gravitational field, where gravitational metrics are restricted to be of the form given by (40). From this point of view the relativistic heat equation is nothing more (or less) then the Einstein equation for the class of metrics given by (40). Therefore, we can speak of phonon gravity as an appropriate designation of the phonon contribution to the thermal Hamiltonian.

The remarkable fact about (42) is that it is the Hamiltonian for zero temperature and therefore is a Hamiltonian for a system at thermal gravitational equilibrium. We thus come to the conclusion that in our case a correspondence can be established between non-equilibrium quantum field systems at zero gravity but at non-zero spatially varying temperature and an equilibrium gravitating system at zero temperature. The correspondence suggests the existence of what may be called the Tolman thermal equivalence principle between non-equilibrium thermal field theories on flat space-times and equilibrium field theories on curved space-times.

In summary we extended the Keldysh method to the steady state non-equilibrium case, elucidated the relation between the quasi-classical Boltzmann equation and two-point functions of the quantum field theory as applied to the thermoelectric effect, derived physically appealing equation on the temperature field non-relativistic case and proposed an equation for the relativistic. Being able to determine temperature distributions inside the materials from the electronic properties of the matter should be quite useful for many practical applications physics.

## Acknowledgements

This work was conceived during a visit to the Theoretical Physics Institute, TU Dresden in fall of 2013. I wish to thank the Institute for the hospitality. I would also like to thank Carsten Timm for introducing me to the subject and for many useful discussions.